\documentclass[11pt]{article}

\usepackage{amsmath,amsfonts,amssymb,amscd,latexsym,mathrsfs,amsthm}

\newtheorem{theorem}{Theorem}

\newtheorem{corollary}[theorem]{Corollary}

\newtheorem{remark}[theorem]{Remark}

\newcommand{\C}{\ensuremath{\mathbb{C}}}

\newcommand{\R}{\ensuremath{\mathbb{R}}}

\def\H{\mathcal H}
\def\d{\mathrm{d}}

\def\Diag{\mathop{\mathsf{Diag}}\nolimits}
\def\Con{{\mathrm{Const.~}}}
\def\1{\mathbf 1}

\begin{document}

\title{Schr\"odinger operators with $n$ positive eigenvalues: \\
an explicit construction involving complex valued potentials}

\author{Serge Richard$\,^1$\footnote{On leave from Universit\'e de Lyon;
Universit\'e Lyon 1; CNRS, UMR5208, Institut Camille Jordan,
43 blvd du 11 novembre 1918, F-69622 Villeurbanne-Cedex, France.
Supported by JSPS Grant-in-Aid for Young Scientists~A no. 26707005.},~~Jun Uchiyama$\,^2$~and Tomio Umeda$\,^3$\footnote{Supported by
the by JSPS Grant-in-Aid for Scientific Research C no. 26400175.}}
\date{}
\maketitle

\vspace{-8mm}

\begin{quote}
\begin{itemize}
\item[$^1$] Graduate school of mathematics,
University of Nagoya,
Chikusa-ku, \\
Nagoya 464-8602,
Japan \\
E-mail: {\tt richard@math.nagoya-u.ac.jp}
\item[$2$] Kyoto Institute of Technology, Sakyo-ku, Kyoto 606-8585, Japan \\
E-mail: {\tt jun-uchi@mbox.kyoto-inet.or.jp}
\item[$3$] Department of Mathematical Sciences, University of Hyogo, Shosha,\\ Himeji 671-2201, Japan \\
E-mail: {\tt umeda@sci.u-hyogo.ac.jp}
\end{itemize}
\end{quote}

\begin{abstract}
An explicit construction is provided for embedding $n$ positive eigenvalues in the spectrum of a
Schr\"odinger operator on the half-line with a Dirichlet boundary condition at the origin. The resulting potential
is of von Neumann-Wigner type, but can be real valued as well as complex valued.
\end{abstract}

\section{Introduction}
Since the seminal paper of von Neumann and Wigner \cite{vNW}, Schr\"odinger operators with embedded positive eigenvalues have
always played a special role in spectral and scattering theory.
In particular, several attempts have been made for generalizing the original result and for studying the stability of the embedded eigenvalues.
It is obviously impossible to mention all references dealing with these questions, but let us mention a few which
are related to our investigations \cite{AU,BD,CHM,L,S}, as well as the books \cite[Chap.~4.4]{EK} and \cite[App.~2, Chap.~XI.8]{RS3}.
Despite these numerous works it seems to the authors of the present note that there is still some room left for discussing the case
of $n$ distinct embedded eigenvalues, especially for complex valued potentials.

Given $n$ positive numbers $\mu_1>\mu_2>\dots>\mu_n>0$, we propose a very simple construction of a von Neumann and Wigner's type potential $V$
such that the corresponding Schr\"odinger operator $-\frac{\d^2}{\d r^2}+V$ on $\R_+ $ with Dirichlet condition at the origin,
admits the eigenvalues $\mu_1^2, \mu_2^2, \dots ,\mu_n^2$.
The potential $V$ as well as the eigenfunctions $v_j$ are explicitly constructed and do not rely on an implicit equation.
In addition, a family of $n$ parameters can still be chosen arbitrarily, asserting once more that these eigenvalues
are stable under suitable modifications of the potential, see for example \cite{AHS,CHM}.
We emphasize that depending on the choice of these parameters, the potential $V$ can be real and thus leads to a self-adjoint operator,
or can be complex valued.
Note however that the leading term of the potential is of the form
\begin{equation}\label{eqform}
V(r)=-\frac{4}{r}\sum_{j=1}^n \mu_j \sin(2\mu_j r) + O(r^{-2}) \qquad \hbox{ as } r\to \infty,
\end{equation}
which implies in particular that the dependence in these parameters take place only in the remainder term at infinity.
In fact, in the main statement below we exhibit the second term of the expansion of the potential and its dependence
on the mentioned parameters.

The possibility of constructing a potential $V$ of the form \eqref{eqform} with $n$ positive eigenvalues has been known for
a long time, see for example \cite{EK,MT,RS3}. Despite this fact, it appears to the authors that an explicit solution has never been provided.
Let us however emphasize that our inspiration came from the paper \cite{MT} and from the book \cite{EK},
in the special case when all mentioned parameters are equal to $1$
(or more precisely when the matrix $A$ of Theorem \ref{construction} is equal to the identity matrix).
Moses and Tuan based their example on the Gel'fand-Levitan theory for inverse problem, see \cite{GL}.
By adapting their idea, one can construct directly a potential having several positive eigenvalues,
but it is not necessary to use the theory of Gel'fand and Levitan.
Let also mention that an alternative construction has been proposed in \cite{S} which leads to the possibility
of embedding a finite or an infinite number of positive eigenvalues, but even in the finite case, the resulting potential
is not of the form of the one we exhibit. In addition, it seems that most if not all previous works were dealing with real valued potentials only.

In the last part of this note, we mention that our main result, obtained for Schr\"odinger operator on $\R_+$ with a Dirichlet condition at the origin, also leads
to a similar result for Schr\"odinger operators on $\R^3$ with the spherically symmetric potential $V(|\cdot|)$. We also show that such a simple
construction can only take place in $\R^3$ and not in any other dimension $\R^d$ if $d$ is different from $1$ or $3$.

Let us finally mention that part of the results of the present note was announced in \cite{U}.


\section{The main result}\label{secmain}

For any $\mu_1>\mu_2>\dots>\mu_n>0$ and any $r\in \R$, let us set
\begin{align*}
s(r)&:={}^t\!\big(\sin(\mu_1 r), \sin(\mu_2 r),\dots,\sin(\mu_n r)\big) \\
c(r)&:={}^t\!\big(\cos(\mu_1 r), \cos(\mu_2 r),\dots,\cos(\mu_n r)\big) \\
g_{ij}(r)&:=\int_0^r \sin(\mu_i\rho)\;\!\sin(\mu_j\rho)\;\!\d \rho,
\end{align*}
where ${}^t\!(\cdot)$ means the transposed vector (column vector).
Note that $g_{ij}=g_{ji}$, and by taking the equality
\begin{equation*}
\sin(\mu_i\rho)\sin(\mu_j\rho)= \frac{1}{2}\Big(\cos\big((\mu_i-\mu_j)\rho\big)-\cos\big((\mu_i+\mu_j)\rho\big)\Big)
\end{equation*}
into account, one also infers that
\begin{equation}\label{matrixG}
g_{ij}(r)=
\begin{cases}
h_{ij}(r) & \hbox{ for } i\neq j \\
\noalign{\vskip 5pt}
\displaystyle{\frac{r}{2}} + h_{ii}(r) & \hbox{ for } i=j
\end{cases}
\end{equation}
with
\begin{equation*}
h_{ij}(r) =
\begin{cases}
\displaystyle{\frac{\sin\big((\mu_i -\mu_j)r\big)}{2 (\mu_i - \mu_j)}
- \frac{\sin\big((\mu_i +\mu_j)r\big)}{2 (\mu_i + \mu_j)} }
  & \hbox{ for }\ i\not=j \\
  \noalign{\vskip 6pt}
\displaystyle{-\frac{\sin(2\mu_i r)}{4 \mu_i}}
   &  \hbox{ for }\ i = j .
\end{cases}
\end{equation*}
Both expressions for $g_{ij}$ will be useful later on.

We also define the $n\times n$ hermitian matrix $G:=\big(g_{ij}\big)_{i,j=1}^n$,
and set $\1_n$ for the $n\times n$ identity matrix.
In the sequel we write $C^\infty\big([0,\infty)\big)$ for smooth functions on $(-\varepsilon,\infty)$, for some $\varepsilon>0$,
but restricted to the subset $[0,\infty)$.
Finally, we write $\C^*$ for $\C\setminus \{ 0 \}$, and use the standard notation $f'$ for the derivative of a function $f$ with respect to its variable.
Our main result then reads\ :

\begin{theorem}  \label{construction}
Let $A=\Diag(a_1, \, \cdots, \, a_n)$ be a diagonal $n\times n$ matrix
with $a_j \in \{z\in \C^*\mid \Re(z)\geq 0\}$. Then, $\big(A+G(r)\big)$ is invertible for any $r\geq 0$,
and by setting
\begin{equation*}
v(r):=-\big(A+G(r)\big)^{-1}s(r)
\qquad \hbox{ and } \qquad
V(r):=2\Big(\sum_{j=1}^n \sin(\mu_j \cdot)\;\!v_j(\cdot)\Big)'(r),
\end{equation*}
the following properties hold:
\begin{enumerate}
\item[(i)] $V\in L^\infty(\R_+)\cap C^{\infty}\big([0,\infty)\big)$ and satisfies for $r\to \infty$
\begin{equation}\label{expansion1}
V(r)= -\frac{4}{r}\sum_{j=1}^n \mu_j \;\!\sin(2\mu_j r)
+ \frac{8}{r^2} \Big(\sum_{j=1}^n a_j \;\!\mu_j \;\!\sin (2\mu_j r)
+ W(r)\Big) + O(r^{-3}),
\end{equation}
with $W$ the real valued function given by
\begin{equation*}
W(r) =  \Big( \sum_{j=1}^n \sin^2 (\mu_j r) \Big)^2
+ 2 \sum_{i,j=1}^n h_{ij}(r)\;\! \mu_i \;\!\sin (\mu_j r)\;\! \cos(\mu_i r).
\end{equation*}
\item[(ii)] For $j\in\{1,\dots,n\}$, the component $v_j$ of $v$ belongs to $C^{\infty}\big([0,\infty)\big)$ and
satisfies $|v_j(r)| \le {\rm Const.} \frac{r}{1 + r^2}$ for any $r\in \R_+$.
In addition, $v_j$ is a solution of $\big(-\frac{\d^2}{\d r^2}+V\big)v_j=\mu_j^2 v_j$, where the Dirichlet realization
of the operator $-\frac{\d^2}{\d r^2}+V$ is considered in $L^2(\R_+)$.
\end{enumerate}
\end{theorem}

\begin{remark}
Let us emphasize that the essential spectrum $[0,\infty)$ of the Dirichlet realization
of $-\frac{\d^2}{\d r^2}+V$, as well as its embedded eigenvalues
$\mu_j^2$, are independent of any choice of the parameters $a_j$,
as long as the invertibility condition of $A + G(r)$ is ensured.
This property is quite remarkable since
the $v_j$'s and hence the potential $V$ depend on $A$, as shown in \eqref{expansion1}.
Note that in this expression, we have emphasized the linear dependence on the parameters $a_j$ in the second order term,
but the remainder term also depend on them.
Note also that $V$ is real-valued, and thus $-\frac{\d^2}{\d r^2}+V$ is self-adjoint, if $a_j>0$ for all $j\in \{1, \dots, n\}$, but
is complex-valued if $\Im(a_j)\neq 0$ for some $j$.
This persistence of the embedded eigenvalues under lower order modifications of the
potential is consistent with the results contained in \cite{AHS,CHM}.
\end{remark}

\begin{remark}
For information and as we shall see in \eqref{def-v}, \eqref{estimate_inv} and
in \eqref{expansion3}, each $v_j$ satisfies the expansions
for $r \to 0$
\begin{equation*}
v_j(r) = - a_j^{-1}\mu_j r + O(r^4),
\end{equation*}
and for $r \to \infty$
\begin{equation*}
v_j(r) = -\frac{2}{\, r \,} \sin (\mu_j r) +
\frac{4}{\, r^2 \,} \big\{ a_j \sin (\mu_j r) + \sum_{l=1}^n h_{jl}(r)\sin(\mu_l r) \big\}
+  O(r^{-3}).
\end{equation*}
In a similar way, the asymptotic expansions of $v_j^{\prime}(r)$ as $r \to \infty$
can be obtained from \eqref{expansion4}, and the asymptotic expansion of
$v_j^{\prime\prime}(r)$ as $r \to \infty$ follows readily from
the relation $v_j^{\prime\prime}= (-\mu_j^2+V) v_j$.
\end{remark}

The following proof is divided into several small pieces. We use the notation $M_n(\C)$ for the set of all $n\times n$ complex matrices,
and the notation $\langle \xi,\zeta\rangle := \sum_{j=1}^n \overline{\xi}_j\zeta_j$ for the usual scalar product in $\C^n$.

\begin{proof}
a) Let us first show the invertibility of $A+G(r)$ for any $r\geq 0$. For $r=0$,
it follows from the definition that $G(0)=0$, and then $A$ is invertible because each $a_j\neq 0$.
Assume now that there exists $\xi \in \C^n\setminus\{0\}$ which belongs to the kernel of $A+G(r)$ for some fixed $r>0$.
Then one has
\begin{equation}\label{eq_contra}
0 = \big\langle \xi, \big(A+G(r)\big)\xi\big\rangle = i\langle \xi,\Im(A)\xi\rangle + \langle \xi, [\Re(A)+ G(r)]\xi\rangle.
\end{equation}
By assumption on $a_j$, it follows that $\Re(A)\geq 0$, while for the second real term one has
\begin{equation*}
\langle \xi,G(r)\xi\rangle = \int_0^r \sum_{i,j=1}^n \overline{\xi_i}\xi_j \sin(\mu_i \rho)\sin(\mu_j \rho)\;\!\d \rho
= \int_0^r \Big|\sum_{i=1}^n \xi_i\sin(\mu_i\rho)\Big|^2\d \rho > 0.
\end{equation*}
One infers from these estimates that the real term in \eqref{eq_contra} can not be equal to $0$, leading thus to a contradiction.

Since $A+G(r)$ is invertible for any $r\geq 0$, one sets
\begin{equation} \label{def-v}
v(r):= -\big(A+G(r)\big)^{-1}s(r)
\end{equation}
and readily infers that $v\in C^\infty\big([0,\infty);\C^n\big)$.

b) Let us observe that for any fixed $r>0$ and by the mean value theorem, there exists $\theta = \theta(r)\in (0,1)$ such that
$g_{ij}(r)= r\sin(\mu_i\theta r)\sin(\mu_j\theta r)$, from which one deduces that
\begin{equation*}
\left|\frac{1}{r^3}g_{ij}(r)\right| =\left|\frac{\sin(\mu_i\theta r)}{\mu_i\theta r}\right|
\;\!\left|\frac{\sin(\mu_j\theta r)}{\mu_j\theta r}\right| \mu_i\mu_j \theta^2 \leq \mu_i\mu_j.
\end{equation*}
Consequently, from this estimate and from the explicit expression provided in \eqref{matrixG} the following properties hold:
\begin{equation*}
A + G(r)
\ = \ \left\{\begin{matrix}
\frac{r}{2}\1_n + O(1) & \hbox{ as } r\to \infty \\
A + O(r^{3}) & \hbox{ as } r\to 0
\end{matrix}\right.
\ = \ \left\{\begin{matrix}
\frac{r}{2}\big(\1_n + O(r^{-1})\big) & \hbox{ as } r\to \infty \\
A + O(r^{3}) & \hbox{ as } r\to 0\;\!,
\end{matrix}\right.
\end{equation*}
and for the inverse of this matrix, one deduces that
\begin{equation}\label{estimate_inv}
\big(A + G(r)\big)^{-1} = \left\{\begin{matrix}
\frac{2}{r}\1_n + O(r^{-2}) & \hbox{ as } r\to \infty \\
A^{-1} + O(r^{3}) & \hbox{ as } r\to 0\;\!.
\end{matrix}\right.
\end{equation}

As a consequence of these estimates, one infers from the definition of $v(r)$ and from the estimate
$|\sin(\mu_i r)|\leq \mu_i r$ that for any $j\in\{1,\dots,n\}$ and any $r\geq 0$
\begin{equation}\label{estimate_V}
|v_j(r)|\leq \Con\!\frac{r}{1+r^2}.
\end{equation}
It thus follows that $v_j\in L^2(\R_+)$.

c) Let us now consider $v'$, the derivative of $v$ with respect to its variable. Since
\begin{equation}  \label{eq_0}
v'(r) = \big(A + G(r)\big)^{-1} G'(r) \big(A + G(r)\big)^{-1} s(r) - \big(A + G(r)\big)^{-1} s'(r),
\end{equation}
one deduces from \eqref{estimate_inv} and from the definition of $G(r)$ that $|v_j'(r)|\leq \Con\!(1+r)^{-1}$, for any $r\geq 0$ and any $j\in \{1,\dots,n\}$.
Note that a simple consequence of this estimate is that $V\in L^\infty(\R_+)$. In addition, the regularity property of $V$ can easily be deduced from the
corresponding properties of the functions $v$ and $s$.

d) Let us now check that the equality $\big(-\frac{\d^2}{\d r^2}+V\big)v=M^2 v$ holds, with $M$ the diagonal $n\times n$ matrix $\Diag(\mu_1, \mu_2,\dots,\mu_n)$. \
In order to compute the expression $-v'' + Vv$, observe that from the initial relation $\big(A + G(r)\big)v(r)=-s(r)$ one infers that
\begin{equation*}
G'(r)v(r) + \big(A + G(r)\big) v'(r) = -M c(r)
\end{equation*}
and that
\begin{equation*}
G''(r)v(r)+ 2G'(r)v'(r) + \big(A + G(r)\big)v''(r) = M^2 s(r),
\end{equation*}
or equivalently that
\begin{equation*}
\big(A + G(r)\big)v''(r) = M^2 s(r) - 2G'(r)v'(r) - G''(r)v(r).
\end{equation*}
From these relations, one then deduces that
\begin{align}
\nonumber &\big(A + G(r)\big)\Big(v''(r)- [Vv](r) + M^2 v(r)\Big) \\
\nonumber = & M^2 s(r) - 2G'(r)v'(r) - G''(r)v(r) - V(r)\big(A + G(r)\big)v(r) \\
\nonumber & \ + M^2\big(A + G(r)\big)v(r) + [G(r),M^2]v(r) \\
\label{eq_1} = & - 2G'(r)v'(r) - G''(r)v(r) + V(r)s(r) +  [G(r),M^2]v(r),
\end{align}
where $[\cdot,\cdot]$ is used for the commutator of two matrices.

On the other hand, observe that $G'(r) = \big(\sin(\mu_i r)\sin(\mu_j r)\big)_{i,j=1}^n = s(r)\;\!{}^t\!s(r)$,
where we have used the identification of $\C^n$ with the matrices $M_{n1}(\C)$.
As a consequence, one infers that
\begin{equation}\label{eq_2}
G'(r)v'(r) = \big\langle s(r),v'(r)\big\rangle \;\! s(r),
\end{equation}
and that
\begin{equation}\label{eq_3}
G''(r)v(r) = \big\langle s(r),v(r)\big\rangle \;\! s'(r) + \big\langle s'(r),v(r)\big\rangle \;\! s(r).
\end{equation}
Let us also observe that $\big([G(r),M^2]\big)_{ij} = (\mu_j^2-\mu_i^2)g_{ij}(r)$. In addition,
by taking the equalities $\mu_j^2\sin(\mu_j \rho) = -[\sin(\mu_j \cdot)''](\rho)$ into account,
one also observes that
\begin{align*}
(\mu_j^2-\mu_i^2)g_{ij}(r)
& = \int_0^r \sin(\mu_i\rho)\;\!\mu_j^2\sin(\mu_j\rho)\;\!\d \rho
- \int_0^r \mu_i^2 \sin(\mu_i\rho)\;\!\sin(\mu_j\rho)\;\!\d \rho \\
& = \Big[\sin(\mu_i \rho)\big(-[\sin(\mu_j \cdot)'](\rho)\big)\Big]_0^r
+ \int_0^r [\sin(\mu_i\cdot)'](\rho)\;\![\sin(\mu_j\cdot)'](\rho)\;\!\d \rho \\
& \quad - \Big[\big(-[\sin(\mu_i \cdot)'](\rho)\big)\sin(\mu_j \rho)\Big]_0^r
- \int_0^r [\sin(\mu_i\cdot)'](\rho)\;\![\sin(\mu_j\cdot)'](\rho)\;\!\d \rho \\
& = \big[-\sin(\mu_i \cdot) \sin(\mu_j \cdot)' + \sin(\mu_i\cdot)'\sin(\mu_j \cdot)\big](r).
\end{align*}
Consequently, one has obtained that
\begin{equation}\label{eq_4}
[G(r),M^2] = -s(r)\;\!{}^t\!s'(r) + s'(r)\;\!{}^t\!s(r).
\end{equation}

By inserting now the equalities \eqref{eq_2}, \eqref{eq_3} and \eqref{eq_4} into \eqref{eq_1} one infers that
\begin{align*}
&\big(A + G(r)\big)\Big(v''(r)- [Vv](r) + M^2 v(r)\Big) \\
= & - 2 \big\langle s(r),v'(r)\big\rangle \;\! s(r)
- \big\langle s(r),v(r)\big\rangle \;\! s'(r) - \big\langle s'(r),v(r)\big\rangle \;\! s(r) + V(r)s(r) \\
& \quad  -\big\langle s'(r),v(r)\big\rangle \;\! s(r) + \big\langle s(r),v(r)\big\rangle \;\!s'(r)\\
= & -2\Big( \big\langle s(r),v'(r)\big\rangle
+ \big\langle s'(r),v(r)\big\rangle \Big) s(r) + V(r)s(r) \\
= & -2\big[\big\langle s(\cdot),v(\cdot)\big\rangle'\big](r)s(r) + V(r)s(r) \\
= & 0
\end{align*}
since $-2 [\big\langle s(\cdot),v(\cdot)\big\rangle'](r) = -2  \Big(\sum_{j=1}^n \sin(\mu_j \cdot)\;\!v_j(\cdot)\Big)'(r) = -V(r)$.
Finally, since $A + G(r)$ is invertible, one infers from the previous computation that
$v''(r)- [Vv](r) + M^2 v(r)=0$, or equivalently that
$-v''(r) + [Vv](r)= M^2 v(r)$, as expected.

e) It has been shown in the point b) that $v_j\in L^2(\R_+)$ for any $j\in \{1,\dots,n\}$, and the equality $v_j(0)=0$ clearly holds.
In addition, it follows from the estimate obtained in c) that $v'_j\in L^2(\R_+)$. Finally, since $V\in L^\infty(\R_+)$, as pointed out in c),
one infers from the equality $-v''=M^2v-Vv$ that $v_j''\in L^2(\R_+)$ as well. Thus, $v_j$ belongs
to the Dirichlet realization of the Laplace operator on $\R_+$, and this concludes the proof of the second statement of the proposition.

f) It only remains to derive the asymptotic expansion \eqref{expansion1}.
For that purpose, let us set $H := \big(h_{ij} \big)_{i,j=1}^n$,
and let us use the notation $\|\cdot \|$ for the norms on $\C^n$ and on $M_n(\C)$.
By taking \eqref{matrixG} into account one gets
\begin{equation*}
A + G(r)= \frac{\,r}{2} \Big(
\1_n + \frac{2}{r}\big( A + H(r) \big)\Big).
\end{equation*}
Since $\| H(r) \| \le C$ for all $r \ge 0$ with a constant $C$
independent of $r$, one deduces that there exists $r_0>0$ such that for any $r\geq r_0$
\begin{equation*}
\frac{2}{r}\| A + H(r) \| < 1/2\  .
\end{equation*}
From the Neumann series, one then infers that
\begin{equation} \label{expansion2}
\big(A + G(r) \big)^{-1} =\frac{2}{r} \1_n - \frac{4}{r^2} \big( A + H(r) \big) + O(r^{-3})
\;\;\; \text{as }r \to \infty,
\end{equation}
and by taking \eqref{expansion2} and \eqref{eq_0} into account,
it follows that
\begin{equation} \label{expansion3}
v(r) = -\frac{2}{\,r}s(r) + \frac{4}{\,r^2} \big\{ A + H(r) \big\} s(r) + O(r^{-3}),
\end{equation}
and that
\begin{equation} \label{expansion4}
v^{\prime}(r) =  -\frac{2}{\,r}Mc(r) +
\frac{4}{\,r^2} \big\{ s(r) \, {}^t\!s(r) \, s(r)  + AMc(r) + H(r) M c(r) \big\} + O(r^{-3}).
\end{equation}
Finally, by putting together these information one obtains
\begin{align*}
V(r) & = 2 \big\langle s(r), v'(r)\big\rangle
+ 2 \big\langle s'(r), v(r)\big\rangle \\
& = 2 \big\langle s(r), v'(r)\big\rangle
+ 2 \big\langle Mc(r), v(r)\big\rangle \\
& = -\frac{4}{r}\langle s(r), Mc(r)\rangle + \frac{8}{r^2}\big\langle s(r),\big\{ s(r) \, {}^t\!s(r) \, s(r)  + AMc(r) + H(r) M c(r) \big\}\big\rangle \\
& \qquad - \frac{4}{r}\langle Mc(r),s(r)\rangle + \frac{8}{r^2}\big\langle Mc(r),  \big\{ A + H(r) \big\} s(r) \big\rangle + O(r^{-3}) \\
& = -\frac{8}{r}\big\langle s(r),Mc(r)\big\rangle
+ \frac{8}{\,r^2}
\big\{
\big\langle s(r), A Mc(r)\big\rangle
+\big\langle Mc(r), A s(r)\big\rangle
\big\}  \\
& \qquad +
\frac{8}{\,r^2}
\big\{ \, \|s(r)\|^4  + \big\langle s(r), H(r) Mc(r) \big\rangle
+ \big\langle Mc(r), H(r) s(r)\big\rangle
\big\} + O(r^{-3}) \\
& = -\frac{4}{r}\sum_{j=1}^n \mu_j \sin(2\mu_j r)
+
\frac{8}{\,r^2} \sum_{j=1}^n a_j \mu_j \sin (2\mu_j r)  + \frac{8}{\,r^2}W(r) +
 O(r^{-3}),
\end{align*}
which gives the expansion \eqref{expansion1}.
\end{proof}

By using the standard relation between the Dirichlet Laplacian on $\R_+$ and the restriction of the Laplace operator $-\Delta$ on $\R^3$ to spherically
symmetric functions (see for example \cite[Sec.~11.3]{AJS}), the previous result easily leads to a similar statement on $\R^3$.

\begin{corollary}
The operator $-\Delta+V(|\cdot|)$, with domain the Sobolev space $\H^2(\R^3)$,
admits $n$ eigenfunctions $u_j$ satisfying
$(-\Delta+V(|\cdot|))u_j = \mu_j^2 u_j$.
These eigenfunction $u_j$ are given by $u_j(x):=v_j(|x|)/|x|$ for any $x\in \R^3$ and with $v_j$ defined in Theorem \ref{construction}.
\end{corollary}

Let us finally show that this construction is valid in $\R^3$ only.
Indeed, if we consider the $d$-dimensional Laplacian acting on
spherically symmetric functions of the form $u_j(x)=a(r) v_j(r)$ with $a,v_j\in C^\infty\big((0,\infty)\big)$, $x\in \R^d\setminus\{0\}$ and $r=|x|$,
then we find that
\begin{align*}
[\Delta u_j] (x)
&= u_j''(r) + \frac{d-1}{r} u_j'(r) \\
&= a(r) v_j''(r) + \big\{ 2 a'(r) + \frac{d-1}{r}a(r) \big\} v_j'(r)
+ \big\{  a''(r) + \frac{d-1}{r}a'(r) \big\} v_j(r).
\end{align*}
In order to use a relation of the form $-v_j''+Vv_j = \mu_j^2v_j$,
it is thus necessary to impose that
\begin{equation*}
2 a'(r) + \frac{d-1}{r}a(r)=0
\qquad \hbox{and}\qquad
a''(r) + \frac{d-1}{r}a'(r) =0.
\end{equation*}
The first equation has the unique solution (up to constants)
given by $a(r) = r^{(-d+1)/2}$, and by substituting this solution into the second equation one obtains
\begin{equation*}
a''(r) + \frac{d-1}{r}a'(r)
= -\frac{(d -1)(d-3)}{4} \;\!r^{-(d +3)/2} = 0
\end{equation*}
which means that $d$ can only be equal to $3$ (the case $d=1$ and $a=const.$ clearly corresponds to even functions on $\R$).


\end{document}